\documentclass[11pt]{article}
\usepackage{amsmath,amssymb}

\usepackage{graphicx}
\usepackage{wrapfig}
\usepackage{float}

\usepackage{hyperref}
\usepackage{cite}

\def\beq{\begin{eqnarray}}
\def\eeq{\end{eqnarray}}

\newcommand{\CA}{{\mathcal{A}}}

\newcommand{\CJ}{{\mathcal{J}}}

\newcommand{\Tr}{\,\mathrm{Tr}\,}            

\newcommand{\be}{\begin{equation}}
\newcommand{\ee}{\end{equation}}
\newcommand{\bea}{\begin{eqnarray}}
\newcommand{\eea}{\end{eqnarray}}
\newcommand{\bg}{\begin{gather}}

\newcommand{\bseq}{\begin{subequations}}
\newcommand{\eseq}{\end{subequations}}

\def\half{\frac{1}{2}}

\def\be{\begin{eqnarray}}
\def\ee{\end{eqnarray}}
\def\lb{\label}

\begin{document}

\title{\textbf{Entropy discrepancy and total derivatives \\
in  trace anomaly }}

\vspace{2cm}
\author{ \textbf{  Amin Faraji Astaneh$^{1}$, Alexander Patrushev$^{2,3}$ and  Sergey N. Solodukhin$^3$ }} 
\date{}
\maketitle
\begin{center}
\hspace{-0mm}
  \emph{$^1$ Department of Physics, Sharif University of Technology,}\\
\emph{P.O. Box 11365-9161, Tehran, Iran\\
 and  School of Particles and Accelerators,\\ Institute for Research in Fundamental Sciences (IPM),}\\
\emph{ P.O. Box 19395-5531, Tehran, Iran}
 \end{center}
 \begin{center}
  \hspace{-0mm}
  \emph{ $^2$ Bogoliubov Laboratory of Theoretical Physics,}\\
  \emph{Joint Institute for Nuclear Research,}\\
  \emph{6 Joliot-Curie, 141980 Dubna, Russia}
\end{center}
\begin{center}
  \hspace{-0mm}
  \emph{ $^3$ Laboratoire de Math\'ematiques et Physique Th\'eorique  CNRS-UMR
7350, }\\
  \emph{F\'ed\'eration Denis Poisson, Universit\'e Fran\c cois-Rabelais Tours,  }\\
  \emph{Parc de Grandmont, 37200 Tours, France}
\end{center}



\begin{abstract}
\noindent { In this note we address the discrepancy found by Hung, Myers and Smolkin between the holographic calculation  of entanglement entropy (using the Jacobson-Myers functional for the holographic minimal surface) and the CFT trace anomaly calculation
if one uses the Wald prescription to compute the entropy in six dimensions. As anticipated in our previous work  \cite{Astaneh:2014sma} the discrepancy originates entirely from a total derivative term present 
in the trace anomaly in six dimensions.
}
\end{abstract}

\vskip 1 cm
\noindent
\rule{7.7 cm}{.5 pt}\\
\noindent 
\noindent
\noindent ~~~ {\footnotesize e-mails: faraji@ipm.ir,   apatrush@gmail.com,  Sergey.Solodukhin@lmpt.univ-tours.fr}

\newpage
    \tableofcontents
\pagebreak

\newpage

\section{ Introduction}

The systematic study of entropy associated to the gravitational action has been initiated by works of Wald \cite{Wald} and was initially motivated by
applications to the entropy of black hole horizons \cite{JM}.  It was realized that the formula for computing the entropy is in a certain correspondence
with the terms in the gravitational action. The usual area law for black hole horizons in the Einstein theory of gravity is necessarily modified as soon as
the action includes terms of higher power in curvature.
The conical singularity method  introduced in \cite{FS} has sharpened this correspondence and provided an efficient algorithm to compute the entropy
in a way which does not a priori require the metric to satisfy any field equations. For the Killing horizons this off-shell method is in a complete agreement with the Wald
prescription although the latter requires the metric to be on-shell, i.e. satisfy some gravitational equations.
This method is purely geometrical. It explores the distributional nature of the conical singularities.
Later it was realized that the conical singularity method has a much wider applicability and can be used very efficiently to compute the entanglement entropy associated to an
arbitrary surface $\Sigma$, not necessarily a black hole horizon. The background metric in these calculations  a priori should not satisfy 
any field equations. Thus, that the conical singularity method is an   off-shell method is  a clear  advantage. 

The applicability of the method became even wider after the formulation of the holographic description of entanglement entropy \cite{Ryu:2006ef}
(and the proofs in \cite{Dima} and  \cite{Lewkowycz:2013nqa}).
The holography has put in the focus the conformal field theories.
In a related development it was studied a relation between the trace anomaly in a 4d CFT and the  logarithmic terms in entanglement entropy.
It was found in \cite{Solodukhin:2008dh} that this is a one-to-one relation which, in particular, involves  the extrinsic geometry of the entangling surface. 
This observation was the first indication of a departure from the Wald entropy. In terms of the distributional geometry of conical singularities it
manifests  in the appearance of the extrinsic curvature contribution in the integrals of the  invariants quadratic in curvature, as was demonstrated in \cite{Fursaev:2013fta}.
Building on these approaches some further generalizations for more general curvature invariants\cite{Dong}, \cite{Camps}  and applications for holographic calculations \cite{Hung:2011xb},  \cite{Myers:2013lva}, \cite{Bhattacharyya:2013gra} have appeared in the literature. 
On the other hand, there have not yet been much progress in understanding the entropy which originates from the invariants which involve derivatives of the curvature.

Among the  numerous results obtained in the recent years that overwhelmingly  confirmed the theoretical predictions there was one observation which has not yet  found its place
in the otherwise harmonic picture. This observation made by Hung, Myers and Smolkin in 2011  \cite{Hung:2011xb} concerns the entropy in $d=6$ conformal field theory.
They have found  that there is a discrepancy between the holographic calculation  of entanglement entropy (using the Jacobson-Myers functional for the holographic minimal surface) and the CFT trace anomaly calculation
if one uses the Wald prescription to compute the entropy. This discrepancy appears in six dimensions and it is apparently due to the $B_3$ conformal charge.
In \cite{Hung:2011xb} there have been given four examples of rather simple six-dimensional spacetimes and four-dimensional entangling surfaces for which this discrepancy appears.
In all these examples the entangling surfaces are minimal surfaces so that the discrepancy can not be due to the extrinsic curvature which vanishes in these examples.
On the other hand, the $O(2)$ symmetry in the transverse subspace is not present for the surfaces considered in \cite{Hung:2011xb}. Thus, it was emphasized that the discrepancy might be due to the lack of this symmetry.

In our recent work \cite{Astaneh:2014sma} we have suggested that, in a more general context, the total derivative terms in the gravitational action might produce some
non-trivial contributions to the entropy. This is yet a more radical deviation from the Wald prescription for the entropy.
In particular, the total derivative terms in the trace anomaly may give rise to some important contributions to the logarithmic terms
in entanglement entropy of a CFT. As an application of this general statement we have suggested that the discrepancy of  Hung, Myers and Smolkin may originate from the total derivative terms present in the trace anomaly in six dimensions and have proposed some ``phenomenological formula'' for the discrepancy. 
In the present note we finalize  this proposal and identify  explicitly the total derivative term in the $d=6$ trace anomaly that is responsible for the discrepancy.

\section{ Regularized metrics and ``phenomenological'' method}
Motivated by examples of \cite{Hung:2011xb} we shall consider the metric of the following general form
\be
&&ds^2=e^{2\sigma(x,r)}[dr^2+r^2d\tau^2]+g_{ij}(x,r,\tau)dx^idx^j ,
\lb{1} \\
&&g_{ij}(x,r,\tau)=h_{ij}(x)+\frac{1}{2} H_{ij}(x)r^2+\tilde{H}^{ab}_{ij}(x)n^an^b r^2+\cdots\ , \nonumber \\
&&\sigma(x,r)=\sigma_0(x)+\half\sigma_2(x)r^2+\cdots \, \, ,\nonumber 
\ee
where $H_{ij}=H^{ab}_{ij}(x)\delta_{ab}$ is the trace part  and $\tilde{H}^{ab}_{ij}=H^{ab}_{ij}-\frac{1}{2}\delta_{ab}H_{ij}$ is the traceless part of $H^{ab}$, and $n^1=\cos(\tau)$, $n^2=\sin(\tau)$.
In (\ref{1})   we deliberately did not include the terms with extrinsic curvature as we want to study closely the case of \cite{Hung:2011xb}.
The entangling surface $\Sigma$ is at $r=0$ in metric (\ref{1}). Applying the replica trick (for a review on this method see \cite{Solodukhin:2011gn}) we change the periodicity of $\tau$
to be from $0$ to $2\pi n$, where $n$ is an integer. In fact in the replica method we continue to non-integer values of $n$.
This introduces a conical singularity at $r=0$. In order to treat this singularity properly we have to regularize the metric. 

\medskip

\noindent {\it Fursaev-Solodukhin (FS) regularization.} One possible regularization is the one introduced in \cite{FS}. It consists in replacing the metric component $g_{rr}$
(\ref{1}) with $e^{2\sigma(x,r)} f_n(r)$, where
\be
f_n(r)=\frac{r^2+b^2n^2}{r^2+b^2}
\lb{2}
\ee
is the regularization function. At the end of the calculation we are supposed to take the limit $b\to 0$. In many known cases this regularization gives  the Wald entropy.
However, with this regularization alone the regularized metric is characterized by the curvature which is everywhere finite but its derivatives may diverge at $r=0$.
Therefore, one should supplement it with some other  regularization.

\medskip

\noindent {\it Generalized (G) regularization.} This regularization is a generalization of the one introduced in \cite{Fursaev:2013fta}. It is based on the observation that
the divergence in the gradients of the curvature is entirely due to the traceless term $\tilde{H}^{ab}_{ij}$ in the metric (\ref{1}).
Therefore, as suggested in \cite{Camps},  one has to regularize this part of the metric by replacing
\be
H^{ab}_{ij}(x)n^an^b r^2\rightarrow \frac{1}{2}H_{ij}(x) r^2+(H^{ab}_{ij}(x)-\frac{1}{2}\delta^{ab}H_{ij}(x))n^an^b r^{2n}\, 
\lb{3}
\ee
in the metric (\ref{1}). In oder to make the derivatives of curvature regular we assume that $n$ is slightly larger than $1$. If the  traceless part of $H^{ab}$ vanishes then metric (\ref{1}) 
possesses the Killing symmetry  and describes a Killing horizon at $r=0$. For this metric the Wald calculation of entropy is applicable and we do not expect
any modifications of this calculation. This explains why we  did not modify the power of $r$ in front of $H_{ij}(x)$ in (\ref{3}). 
We stress that regularization (\ref{3}) should be used in addition to the regularization with the function $f_n(r)$ (\ref{2}).
This generalized regularization was applied in \cite{Astaneh:2014sma} to the analysis of the contribution of some total derivative terms to the entropy.

Consider now a curvature invariant ${\cal J}$ which may include any function  of curvature and its derivatives. Comparing 
the integrals of ${\cal J}$ in these two regularizations we see that  their difference should vanish provided the traceless part $\tilde{H}^{ab}_{ij}$ vanishes.
Therefore this difference is a function of $\tilde{H}^{ab}_{ij}$ only. To leading order, when only quadratic combinations are taken into account we have that
\be
\left[\int_{{\cal M}_n}{\cal J}\right]_{FS}-\left[\int_{{\cal M}_n}{\cal J}\right]_{G}=(1-n)\int_\Sigma (\alpha \Tr \tilde{H}^{ab}\Tr \tilde{H}^{ab}+\beta  \tilde{H}^{ab}_{ij} \tilde{H}^{ab,ij})\, ,
\lb{41}
\ee
where $\Tr \tilde{H}^{ab}=h^{ij}\tilde{H}^{ab}_{ij}$.

Provided the FS regularization produces the Wald entropy the difference (\ref{41}) gives the desired discrepancy. In the case considered in \cite{Hung:2011xb} 
the invariant ${\cal J}={\cal A}$ is the $d=6$ trace anomaly. In a ``phenomenological'' approach taken in \cite{Astaneh:2014sma}
one can determine the unknown constants $\alpha$ and $\beta$ by making (\ref{41})  consistent with the examples considered in \cite{Hung:2011xb}.
In fact, only two examples of \cite{Hung:2011xb} are sufficient to fix these constants. The expression obtained in  \cite{Astaneh:2014sma} reads
\be
\left[\int_{{\cal M}_n}{\cal A}\right]_{FS}-\left[\int_{{\cal M}_n}{\cal A}\right]_{G}=4\pi(1-n)B_3\int_\Sigma ( \Tr \tilde{H}^{ab}\Tr \tilde{H}^{ab}-4\tilde{H}^{ab}_{ij} \tilde{H}^{ab,ij})\, ,
\lb{4}
\ee
where $B_3$ is the conformal charge which corresponds to invariant $I_3$ in the conformal anomaly. 
It can be rewritten in terms of the doubly traceless tensor $\hat{H}^{ab}_{ij}=\tilde{H}^{ab}_{ij}-\frac{1}{4}h_{ij}\Tr \tilde{H}^{ab}$ as follows
\be
\left[\int_{{\cal M}_n}{\cal A}\right]_{FS}-\left[\int_{{\cal M}_n}{\cal A}\right]_{G}=16\pi(n-1)B_3\int_\Sigma\hat{H}^{ab}_{ij} \hat{H}^{ab,ij}\, .
\lb{4-1}
\ee
This formula is  equivalent to the Hung-Myers-Smolkin expression (equation (5.35) in \cite{Hung:2011xb}) written in terms of the Weyl tensor. Here (and in \cite{Astaneh:2014sma}) we derive this formula in two steps:
first, comparing the two regularization we  conclude that the entropy difference is due to the traceless part of $H^{ab}$ that allowed us to reduce the possible contributions to only two terms (\ref{41}). In the second step, in order to fix the unknown constants $\alpha$ and $\beta$ we have used the values for the entropy discrepancy provided by any two independent   examples considered in \cite{Hung:2011xb}.

\section{ Conformal invariants in six dimensions}
In a generic conformal field theory in $d=6$ the trace anomaly, modulo the total derivatives,  is a combination of four different terms 
\be
<T>={\cal A}=AE_6+B_1I_1+B_2I_2+B_3I_3+(total \ \  derivatives)
\lb{5-1}
\ee
where $E_6$ is the Euler density in $d=6$ and, using notations of \cite{Bastianelli:2000hi}, we have 
\be
&&I_1=W_{\alpha\mu\nu\beta}W^{\mu\sigma\rho\nu}W_{\sigma\ \ \ \rho}^{\ \alpha\beta}\, ,\nonumber \\
&&I_2=W_{\alpha\beta}^{\ \ \mu\nu}W_{\mu\nu}^{\ \ \sigma\rho}W_{\sigma\rho}^{\ \ \alpha\beta}\, , \nonumber \\
&&I_3=W_{\mu\alpha\beta\gamma}\Box W^{\mu\alpha\beta\gamma}+W_{\mu\alpha\beta\gamma} (4R^\mu_\nu-\frac{6}{5}R\delta^\mu_\nu)W^{\nu\alpha\beta\gamma}\, ,
\lb{5}
\ee
where $W_{\alpha\beta\mu\nu}$ is the Weyl tensor.
Notice that, contrary to \cite{Bastianelli:2000hi} we did not include any total derivatives in $I_3$. They will be treated separately later.
In fact we shall also introduce a slightly different version of invariant $I_3$,
\be
I'_3=-\nabla_\sigma W_{\mu\alpha\beta\gamma}\nabla^\sigma W^{\mu\alpha\beta\gamma}+W_{\mu\alpha\beta\gamma} (4R^\mu_\nu-\frac{6}{5}R\delta^\mu_\nu)W^{\nu\alpha\beta\gamma}\, .
\lb{6}
\ee
The difference between these two versions is a total derivative,
\be
I_3-I'_3=\frac{1}{2}\Box W^2\, ,
\lb{7}
\ee
where $W^2=W_{\alpha\beta\mu\nu}W^{\alpha\beta\mu\nu}$ is square of the Weyl tensor.

We  notice  that, as we have checked, for the metric (\ref{1})
there is no difference between the FS and the generalized  regularizations for these two invariants,
\be
\left[\int_{{\cal M}_n} I_k\right]_{FS}=\left[\int_{{\cal M}_n}I_k\right]_{G}\, , \ \ k=1,\  2
\lb{8}
\ee
In particular for the examples considered in \cite{Hung:2011xb} we obtain the values for these invariants completely consistent with those calculated in \cite{Hung:2011xb}
using the Wald method.

\section{Invariants $I_3$ and $I'_3$}
For these invariants the situation is more interesting. What we observe can be summarized as follows.

\medskip

\noindent 1. For invariant $I_3$ there is no difference (up to terms quadratic in $(n-1)$) between the two regularizations\footnote{The divergence of the derivatives of curvature at $r=0$ in  the FS regularization,  that we discussed above, shows up in terms quadratic in $(n-1)^2$.} ,
\be
 \left[\int_{{\cal M}_n} I_3\right]_{FS}=\left[\int_{{\cal M}_n}I_3\right]_{G}+O(n-1)^2\, . \ \ 
\lb{9}
\ee
Thus, the corresponding entropy (obtained using the generalized regularization) agrees with the Wald entropy.

\medskip

\noindent 2. For invariant $I'_3$ the two regularizations differ and the difference is
\be
 \left[\int_{{\cal M}_n} I'_3\right]_{FS}-\left[\int_{{\cal M}_n}I'_3\right]_{G}=4\pi(1-n)\int_\Sigma ( \Tr \tilde{H}^{ab}\Tr \tilde{H}^{ab}-4\tilde{H}^{ab}_{ij}\tilde{H}^{ab,ij})\, . \ \ 
\lb{10}
\ee
This difference  reproduces exactly  the   previous ``phenomenological'' formula (\ref{4}). Clearly, the difference is due to the total derivative term (\ref{7}). This can be independently verified and we indeed get
\be
 \left[\int_{{\cal M}_n} \frac{1}{2}\Box W^2 \right]_{G}=4\pi(1-n)\int_\Sigma ( \Tr \tilde{H}^{ab}\Tr \tilde{H}^{ab}-4\tilde{H}^{ab}_{ij}\tilde{H}^{ab,ij})\, . \ \ 
\lb{11}
\ee
In the FS regularization the integral of the total derivative (\ref{7}) vanishes, as expected.

Let us make some remarks on  the procedure that has led to (\ref{11}). As it was explained in \cite{Astaneh:2014sma}, this procedure is the following.
First we take a disk of finite radius $r_0$ in the plane $(r,\tau)$. For this disk we evaluate the integral of the total derivative term (\ref{7}) for the metric (\ref{1}) regularized according to the generalized regularization procedure.
The integral may include  a boundary term at $r=0$. We however observe that $r\partial_r W^2\sim r^{2n-2}$ vanishes provided that $n$ is slightly above $1$.
Then there remains only a boundary term at $r=r_0$. This boundary term we  first expand in powers of $(n-1)$ and then we take the limit $b\to 0$. The result is expression (\ref{11}).
On the other hand, the calculation in (\ref{10}) is done along the same lines as in \cite{Fursaev:2013fta}: we first do the bulk integration in the radial direction (by replacing $r=bx$ this is integration over $x$ from $0$ to $\infty$), then integrate over $\tau$ and finally take the limit $b\to 0$. That these two calculations (\ref{10}) and (\ref{11})
agree is a nice consistency check.

Equation  (\ref{11}), thus,  reproduces the Hung-Myers-Smolkin discrepancy and clearly indicates that it originates from the total derivative term (\ref{7}).

\section{Some other  total derivative terms}
The calculation above for (\ref{11}) can be repeated for the other total derivative terms.
Here we give some examples,
\be
&&\int_{{\cal M}_n} \Box (R_{\alpha\beta\mu\nu}R^{\alpha\beta\mu\nu})=32\pi(n-1)\int_\Sigma \tilde{H}^{ab}_{ij} \tilde{H}^{ab,ij}+O(n-1)^2\, ,\nonumber \\
&&\int_{{\cal M}_n} \Box (R_{\mu\nu}R^{\mu\nu})=8\pi(n-1)\int_\Sigma \Tr \tilde{H}^{ab}\Tr \tilde{H}^{ab}+O(n-1)^2\, ,\nonumber \\
&&\int_{{\cal M}_n} \Box R^2=O(n-1)^2\, .
\lb{12}
\ee
These are the integrals in the generalized regularization. The FS regularization gives the vanishing results as expected.

\section{Holographic anomaly in $d=6$}
The holographic calculation in \cite{Hung:2011xb} was done in the case of the bulk gravitational action for which the corresponding conformal charges $A$, $B_1$, $B_2$ and $B_3$
are all different. For this case the partition of the total derivative terms among the conformal invariants that correspond to these charges is not known to us.
However, the complete set of total derivatives is available in the case when the $7d$ bulk gravitational action is the Einstein action with a negative cosmological constant, see \cite{Henningson:1998gx} and \cite{Bastianelli:2000hi},
\be
{\cal A}_{hol}=-\frac{L^5}{64\ell_p^5}\big(\half RR_{\mu\nu}R^{\mu\nu}-\frac{3}{50}R^3-R_{\mu\nu\rho\sigma}R^{\mu\rho}R^{\nu\sigma}+
\frac{1}{5}R^{\mu\nu}\nabla_\mu\nabla_\nu R-\half R^{\mu\nu}\Box R_{\mu\nu}+\frac{1}{20}R\Box R\big)\, .
\lb{13-0}
\ee
In this case all conformal charges are proportional to each other,
\be
{\cal A}_{hol}=
\frac{3L^5}{2^5 7!l_p^5}(-\frac{35}{2}E_6-1680I_1-420I_2+140(I_3-C_5)+420C_3-504C_4-84C_6+560C_7)\, ,
\lb{13}
\ee
where $C_5=\frac{1}{2}\Box W^2$ and the other total derivatives $C_k$ are defined in \cite{Bastianelli:2000hi}.
In this case we have $B_3=\frac{L^5}{384 l^5_p}$ for the conformal charge which corresponds to the invariant $I_3$.
In what follows we shall drop the factor $\frac{ L^5}{l^5_p}$ for simplicity.
 That the terms $C_5$ and $I_3$ appear in the combination
$I_3-C_5=I'_3$ is a good sign. In order to complete this analysis we would need to calculate the corresponding contributions which come from the other
terms $C_k$ in (\ref{13}).

The total derivatives  $C_k$ are of the following general  form
\be
\nabla_\mu J^\mu_{(k)}=\nabla_\mu\nabla_\nu C^{\mu\nu}_{(k)}\, ,
\lb{14}
\ee
where $C^{\mu\nu}_{(k)}$ are the following symmetric tensors 
\be
&&C^{\mu\nu}_{(3)}=g^{\mu\nu}(\frac{1}{2}R_{\alpha\beta}^2-\frac{1}{12}R^2)\, ,\nonumber \\
&&C^{\mu\nu}_{(4)}=R^{\mu\sigma}R^\nu_{\ \sigma}-\frac{2}{3}RR^{\mu\nu}+\frac{5}{36}g^{\mu\nu}R^2\, , \nonumber \\
&&C^{\mu\nu}_{(5)}=g^{\mu\nu}(-\frac{1}{2}R_{\alpha\beta}^2+\frac{1}{2}R^2_{\alpha\beta\sigma\rho}+\frac{1}{20}R^2)\, ,\nonumber \\
&&C^{\mu\nu}_{(6)}=RR^{\mu\nu}-R^{\mu\sigma}R^\nu_{\ \sigma}-\frac{1}{4}g^{\mu\nu}R^2\nonumber \\
&&C^{\mu\nu}_{(7)}=R^{\mu\alpha\beta \nu}R_{\alpha\beta}-\frac{5}{4}R^{\mu\sigma}R^\nu_{\ \sigma}+\frac{3}{4}RR^{\mu\nu}+
g^{\mu\nu}(\frac{1}{2}R_{\alpha\beta}^2+\frac{1}{8}R^2_{\alpha\beta\sigma\rho}-\frac{3}{16}R^2)\, .
\lb{15}
\ee

Before proceeding with the computation of the corresponding entropy let us re-write the anomaly (\ref{13-0}), (\ref{13}) in a form without second derivatives of the curvature,
\be
{\cal A}'_{hol}=-\frac{ 1}{64}\big(\frac{1}{2}RR_{\mu\nu}^2-\frac{3}{50}R^3-R_{\mu\alpha\nu\beta}R^{\mu\nu}R^{\alpha\beta}-\frac{3}{20}(\nabla_\lambda R)^2+\frac{1}{2}(\nabla_\lambda R_{\mu\nu})^2\big)\, .
\lb{16}
\ee
The difference between the two forms is a total derivative,
\be
{{\cal A}}'_{hol}={\cal{A}}_{hol}+\frac{3}{320}(C_4+C_6)-\frac{1}{128}C_3\, .
\lb{17}
\ee
Similarly, we shall introduce two forms of anomalies constructed from the conformal invariants,
\be
&&{{\cal A}}_{BI}=\frac{3}{2^5\,  7!}(-\frac{35}{2}E_6-1680I_1-420I_2+140I_3)\, , \nonumber \\
&&{\cal A}'_{BI}=\frac{3}{2^5\,  7!}(-\frac{35}{2}E_6-1680I_1-420I_2+140I'_3)\, , \nonumber \\
&&  {\cal A}'_{BI}={\cal A}_{BI}-\frac{1}{384}C_5\, .
\lb{18}
\ee
It is  convenient for further computations to use the following explicit form of $\CA'_{BI}$
\be
\begin{split}
\CA'_{BI}&=\frac{1}{384}\big(-3RR_{\mu\nu}^2+\frac{9}{25}R^3+4R_{\mu\nu}R^{\nu\alpha}R_\alpha^\mu+2R^{\mu\nu}R^{\alpha\beta}R_{\mu\alpha\nu\beta}-2R^{\mu\nu}R_{\mu\alpha\beta\gamma}R_{\nu}\,^{\alpha\beta\gamma}\\
&+4R_\mu\,^\alpha\,_\nu\,^\beta R^{\mu\rho\nu\sigma}R_{\rho\alpha\sigma\beta}+R_{\alpha\beta}\,^{\mu\nu}R_{\mu\nu}^{\rho\sigma}R_{\rho\sigma}\,^{\alpha\beta}-(\nabla_\lambda  W_{\mu\alpha\nu\beta})^2\big)\, ,
\end{split}
\ee
where in six dimensions
\be
(\nabla_\lambda W_{\mu\alpha\nu\beta})^2=(\nabla_\lambda R_{\mu\alpha\nu\beta})^2-(\nabla_\lambda R_{\mu\nu})^2+\frac{1}{10}(\nabla_\lambda R)^2\, .
\ee

Now, the important relation between the two forms of the anomaly, the holographic and in terms of the conformal invariants, is
\be
{\cal A}'_{hol}={\cal A}'_{BI}+\frac{1}{128}(C_6+\frac{4}{3}C_7) \, .
\lb{19}
\ee
We suggest that namely ${\cal A}'_{hol}$ produces the entropy  that explains the holographic entropy.

\section{Test metric}
Consider a test regularized metric, advantage of which is that it is rather simple for the computation of the respective curvature invariants,
\be
&&ds^2=f_n(r)dr^2+r^2d\phi^2+F_1(r,\phi)dx_1^2+F_2(r,\phi)dx_2^2+dx_3^2+dx_4^2\, ,\lb{20}\\
&&F_1(r,\phi)=1+H_{1}(\cos^2\phi-\sin^2\phi)r^{2n}\, , \nonumber \\
&&F_2(r,\phi)=1+H_{2}(\cos^2\phi-\sin^2\phi)r^{2n}\, .\nonumber
\ee
For this metric we have that
\be
\Tr \tilde{H}^{ab}\Tr \tilde{H}^{ab}=2(H_1+H_2)^2\, , \ \ \Tr(\tilde{H}^{ab}\tilde{H}^{ab})=2(H_1^2+H_2^2)
\lb{21}
\ee
and the surface $\Sigma$ has flat metric $dx_1^2+dx^2_2+dx_3^2+dx_4^2$.
Using our regularization  we compute the entropy which corresponds to the total derivatives for this particular metric and find
\be
&&\int_{{\cal M}_n} \Box R^2=0\, ,\nonumber \\
&&\int_{{\cal M}_n} \Box R^2_{\alpha\beta}=16\pi (H_1+H_2)^2(n-1)\, ,\nonumber \\
&&\int_{{\cal M}_n} \Box R^2_{\alpha\beta\mu\nu}=64\pi(H_1^2+H_2^2)(n-1)\, ,\nonumber \\
&&\int_{{\cal M}_n}  \nabla_\mu\nabla_\nu (RR^{\mu\nu})=16\pi (H_1+H_2)^2(n-1)\, ,\nonumber \\
&&\int_{{\cal M}_n} \nabla_\mu\nabla_\nu (R^{\mu\sigma}R_{\sigma}^{\ \nu})=16\pi (H_1+H_2)^2(n-1)\, , \nonumber \\
&&\int_{{\cal M}_n} \nabla_\mu\nabla_\nu (R^{\mu\alpha\beta\nu}R_{\alpha\beta})=-8\pi (H_1^2+H_2^2)(n-1)\, .
\lb{22}
\ee
With these results we can compute the entropy for the total derivative terms $C_k$,
\be
&&\int_{{\cal M}_n}C_6=0\, , \ \ \int_{{\cal M}_n} C_7=0\, , \nonumber \\
&&\int_{{\cal M}_n}C_5=\int_\Sigma (32\pi (H_1^2+H_2^2)-8\pi (H_1+H_2)^2)(n-1)\, , \nonumber \\
&&\int_{{\cal M}_n}C_3=8\pi\int_\Sigma (H_1+H_2)^2 (n-1)\, , \ \ \int_{{\cal M}_n} C_4=\frac{16}{3}\pi \int_\Sigma (H_1+H_2)^2 (n-1)\, .
\lb{23}
\ee
These findings in particular indicate that the entropy due to the total derivatives $C_6$ and $C_7$ vanishes in equation (\ref{19}) and for the entropy we have an equality
\be
S[{\cal A}'_{hol}]=S[{\cal A}'_{BI}]=-\frac{\pi}{80}\int_\Sigma (8(H_1^2+H_2^2)-(H_1+H_2)^2)\, .
\lb{24}
\ee

{\begin{center}
\begin{table}[H]
\caption{Entropy for various terms in the \textbf{l.h.s} of \eqref{19}. }
\hspace{-1.1cm}
\begin{tabular}{| c | c | c | c | }
  \hline                       
Object ($\CJ$) & Wald's entropy ($S_W[\CJ]$) & Discrepancy ($S_D[\CJ]$) & Total entropy ($S[\CJ]$) \\
  \hline
 \hline
  $R^3$ & $0$ & $0$ & $0$ \\
\hline
$RR_{\mu\nu}^2$ & $8\pi\int_\Sigma(H_1+H_2)^2$ & $0$& $8\pi\int_\Sigma(H_1+H_2)^2$\\
\hline
$R_{\mu\nu\rho\sigma}R^{\mu\rho}R^{\nu\sigma}$ & $-4\pi\int_\Sigma(H_1+H_2)^2$ & $0$  & $-4\pi\int_\Sigma(H_1+H_2)^2$\\
\hline
$(\nabla_\alpha R)^2$ & $-64\pi\int_\Sigma(H_1^2+H_2^2-H_1H_2)$ & $0$ & $-64\pi\int_\Sigma(H_1^2+H_2^2-H_1H_2)$\\
\hline
$(\nabla_\alpha R_{\mu\nu})^2$ & $-16\pi\int_\Sigma(H_1^2+H_2^2)$ & $-8\pi\int_\Sigma(H_1+H_2)^2$ & $-8\pi\int_\Sigma\left(3(H_1^2+H_2^2)+2H_1H_2\right)$\\
\hline
$\CA'_{hol}$ & $-\frac{\pi}{20}\int_\Sigma ((H_1+H_2)^2+2(H_1^2+H_2^2))$ & $\frac{\pi}{16}\int_\Sigma (H_1+H_2)^2$ & $\frac{\pi}{80}\int_\Sigma(-8(H_1^2+H_2^2)+ (H_1+H_2)^2)$\\
\hline
\end{tabular}
\end{table}
\end{center}


{\begin{center}
\begin{table}[H]
\caption{Entropy for  various terms in the \textbf{r.h.s} of \eqref{19}. }
\hspace{-1.6cm}
\begin{tabular}{| c | c | c | c | }
  \hline                       
Object ($\CJ$) & Wald's entropy ($S_W[\CJ]$) & Discrepancy ($S_D[\CJ]$) & Total entropy ($S[\CJ]$) \\
  \hline
 \hline
  $R^3$ & $0$ & $0$ & $0$ \\
\hline
$RR_{\mu\nu}^2$ & $8\pi\int_\Sigma(H_1+H_2)^2$ & $0$& $8\pi\int_\Sigma(H_1+H_2)^2$\\
\hline
$R_{\mu\nu\rho\sigma}R^{\mu\rho}R^{\nu\sigma}$ & $-4\pi\int_\Sigma(H_1+H_2)^2$ & $0$  & $-4\pi\int_\Sigma(H_1+H_2)^2$\\
\hline
$R_{\mu\nu}R^{\nu\alpha}R_\alpha^\mu$ & $12\pi\int_\Sigma(H_1+H_2)^2$ & $0$ & $12\pi\int_\Sigma(H_1+H_2)^2$\\
\hline
$R^{\mu\nu}R_{\mu\alpha\beta\gamma}R_{\nu}\,^{\alpha\beta\gamma}$ & $8\pi\int_\Sigma(H_1^2+H_2^2)$ & $0$ & $8\pi\int_\Sigma(H_1^2+H_2^2)$\\
\hline
$R_\mu\,^\alpha\,_\nu\,^\beta R^{\mu\rho\nu\sigma}R_{\rho\alpha\sigma\beta}$ & $-12\pi\int_\Sigma(H_1^2+H_2^2)$ & $0$ & $-12\pi\int_\Sigma(H_1^2+H_2^2)$\\
\hline 
$R_{\alpha\beta}\,^{\mu\nu}R_{\mu\nu}\,^{\rho\sigma}R_{\rho\sigma}\,^{\alpha\beta}$ & $0$ & $0$ & $0$\\
\hline
$(\nabla_\alpha R)^2$ & $-64\pi\int_\Sigma(H_1^2+H_2^2-H_1H_2)$ & $0$ & $-64\pi\int_\Sigma(H_1^2+H_2^2-H_1H_2)$\\
\hline
$(\nabla_\alpha R_{\mu\nu})^2$ & $-16\pi\int_\Sigma(H_1^2+H_2^2)$ & $-8\pi\int_\Sigma(H_1+H_2)^2$ & $-8\pi\int_\Sigma\left(3(H_1^2+H_2^2)+2H_1H_2\right)$\\
\hline
$(\nabla_\alpha R_{\mu\nu\rho\sigma})^2$ & $0$ & $-32\pi\int_\Sigma (H_1^2+H_2^2)$ & $-32\pi\int_\Sigma (H_1^2+H_2^2)$\\
\hline
$C_5$ & $0$ & $8\pi\int_\Sigma ((H_1+H_2)^2$ & $8\pi\int_\Sigma ((H_1+H_2)^2$\\
$$ & $$ & $-4(H_1^2+H_2^2))$ & $-4(H_1^2+H_2^2))$\\
\hline
$C_6$ & $-8\pi\int_\Sigma (H_1+H_2)^2$ & $8\pi\int_\Sigma (H_1+H_2)^2$ & $0$\\
  \hline  
$C_7$ & $2\pi\int_\Sigma (3(H_1^2+H_2^2)-2H_1H_2)$ & $-2\pi\int_\Sigma (3(H_1^2+H_2^2)-2H_1H_2)$ & $0$\\
\hline
$\CA_{BI}$ & $\frac{\pi}{60}\int_\Sigma (2(H_1+H_2)^2$ & $0$ & $\frac{\pi}{60}\int_\Sigma (2(H_1+H_2)^2$\\
$$ & $-11(H_1^2+H_2^2))$ & $$ & $ -11(H_1^2+H_2^2))$\\
\hline
$\CA'_{BI}$ & $\frac{\pi}{60}\int_\Sigma (2(H_1+H_2)^2$ & $\frac{\pi}{48}\int_\Sigma (-(H_1+H_2)^2$ & $\frac{\pi}{80}\int_\Sigma (-8(H_1^2+H_2^2)$\\
$$ & $-11(H_1^2+H_2^2))$ & $+4(H_1^2+H_2^2))$ & $ +(H_1+H_2)^2)$\\
\hline
$\CA'_{BI}$ & $-\frac{\pi}{20}\int_\Sigma ((H_1+H_2)^2$ & $\frac{\pi}{16}\int_\Sigma (H_1+H_2)^2$ & $\frac{\pi}{80}\int_\Sigma(-8(H_1^2+H_2^2)$\\
$+\frac{1}{128}(C_6+\frac{4}{3}C_7)$ & $+2(H_1^2+H_2^2))$ & $$ & $+ (H_1+H_2)^2)$\\
\hline
\end{tabular}
\end{table}
\end{center}}

This is precisely the holographic entropy computed in \cite{Hung:2011xb}, see also \cite{Huang:2015zua}\footnote{There is a factor of 2 mismatch with \cite{Huang:2015zua}, they use a different normalization for the holographic entropy.}. This confirms our proposal that namely the conformal anomaly in the form ${\cal A}'_{hol}$, that contains no second derivatives of the curvature,
reproduces correctly  the holographic entropy. The difference between the entropy for the two forms of the conformal anomaly, as is seen from (\ref{17}) and (\ref{23}), is
\be
\int_{{\cal M}_n}{\cal A}'_{hol}=\int_{{\cal M}_n} {\cal A}_{hol}-\frac{\pi}{80}\int_\Sigma(H_1+H_2)^2 (n-1)\, .
\lb{25}
\ee
The contribution of each individual term in the anomaly to the entropy and the respective discrepancy are presented in Tables 1 and 2.

We should say that we use here the two different definitions of  ``discrepancy''.
The discrepancy computed in Table 2  is defined as the difference between the total entropy and the Wald entropy computed for the same geometric invariant ${\cal J}$, $S_D({\cal J})=S({\cal J})-S_W({\cal J})$. 
On the other hand, the Hung-Myers-Smolkin discrepancy is, by definition, the difference between  the holographic entropy (identified according to our proposal with the total entropy for the anomaly ${\cal A}'_{hol}$) and the Wald entropy for the anomaly ${\cal A}_{BI}$. These two geometric invariants, ${\cal A}'_{hol}$ and ${\cal A}_{BI}$,  differ by a total derivative term.  
Using our proposal for the holographic entropy and the Table 2 we find that
\be
S_{HMS}=S({\cal A}'_{hol})-S_W({\cal A}_{BI})=S({\cal A}'_{BI})-S_W({\cal A}_{BI})=S_D({\cal A}_{BI})-\frac{1}{384}S_D(C_5)= -\frac{1}{384}S(C_5)\, .
\lb{25-1}
\ee
Thus,  the Hung-Myers-Smolkin discrepancy originates from the entropy of the total derivative term $C_5$, as we anticipated.

It is interesting to compare the Wald entropy and the discrepancy on both sides of  equation (\ref{19}). This is also done in Tables 1 and 2.
The Wald entropy for the total derivatives can be non-zero as was shown in \cite{Astaneh:2015tea}. We used the formulas derived in \cite{Astaneh:2015tea} when calculated
the Wald entropy for the total derivative terms $C_5$, $C_6$ and $C_7$ in Table 2.
We see that both the Wald entropy and the discrepancy are identical on both sides of (\ref{19}). 

We notice that our results for the  total entropy  are consistent with those obtained in \cite{Miao:2014nxa}, \cite{Huang:2015zua}. The agreement\footnote{The criticism  of our approach  in \cite{Huang:2015zua}   is groundless. It is based on the confusion with using the form for the conformal anomaly with the second derivatives of the curvature (${\cal A}_{hol}$)  rather than the one with the gradients (${\cal A}'_{hol}$), the latter form is the correct one as we show in this note.} is quite surprising taking the fact that they use a completely different regularization
in which the total derivatives apparently do not make a contribution to the entropy. On the other hand, it seems impossible ignoring, as in \cite{Miao:2014nxa}, \cite{Huang:2015zua}, the contribution due to the total derivatives to achieve
the same balance between the Wald entropy and the discrepancy on both sides of an identity such as (\ref{19})\footnote{In fact, we have performed a check for the entropy of particular terms entering the holographic anomaly by using  the same regularization as in  \cite{Huang:2015zua} (see their equation (23)). We have found that, contrary to what is claimed in \cite{Huang:2015zua}, the entropy due to the total derivatives is non-vanishing in this regularization. In particular, we have found a non-vanishing result for the combination of total derivatives which appears in eq.(13) of \cite{Huang:2015zua}, contrary to what they claim in eq.(32).}.

\section{Conclusions}
In this note we have completed our earlier proposal made in \cite{Astaneh:2014sma} that the entropy discrepancy observed in \cite{Hung:2011xb} originates from the total derivative terms in the trace anomaly. We have suggested a prescription that  the terms with derivatives of the curvature should be written in the form quadratic in first derivatives rather than containing the second derivatives of the metric in order to reproduce the holographic entropy correctly. We gave a detailed and complete treatment for the entropy of all terms entering in the holographic anomaly (${\cal A}'_{hol}$) and show that the discrepancy in this case originates from the terms that contain gradients of the curvature. 
Together with the resolution of the discrepancy in the entropy of the Maxwell fields, discussed in our previous work  \cite{Astaneh:2014sma} (where we refer to an earlier relevant work of Ch. Eling), our finding  in this note is the second manifestation of the important role of the total derivative terms in the trace anomaly for the logarithmic terms in
entanglement entropy.

\bigskip

\noindent {\bf Acknowledgements:} 
A.P. would like to thank  Laboratoire de Math\'ematiques et Physique Th\'eorique (LMPT) for kind hospitality.
A.P. acknowledges the support from the Metchnikov's post-doctoral program of the Embassy of France  in Russia and the support from the RFBR grant 13-02-00950.

\end{document}